\documentclass[twocolumn,showpacs,preprintnumbers,amsmath,amssymb,aps,prl,superscriptaddress]{revtex4}
\usepackage{graphicx}% Include figure files
\usepackage{dcolumn}% Align table columns on decimal point
\usepackage{bm}% bold math

\begin{document}

\preprint{}

\title{Measurement of the Analyzing Power $A_N$ in $pp$ Elastc Scattering
in the CNI Region with a Polarized Atomic Hydrogen Gas Jet Target}

% author list in APS format for use with revtex4 i.e. for PRL, PRC etc.

\affiliation{Brookhaven National Laboratory, Upton, NY 11973, USA}
\affiliation{RIKEN BNL Research Center, Upton, NY 11973, USA}
\affiliation{Indiana University Cyclotron Facility, Bloomington, IN 47408, USA}
\affiliation{Institute for Theoretical and Experimental Physics (ITEP), 117259 Moscow, Russia}
\affiliation{Kyoto University, Sakyo-ku, Kyoto 606-8502, Japan}
\affiliation{Stony Brook University, Stony Brook, NY 11794, USA}
\affiliation{University of Wisconsin, Madison, WI 53706, USA}
\affiliation{Yale University, New Haven, CT 06520, USA}
% Authors 1 per line follow by affiliation
\author{H.~Okada}\affiliation{Kyoto University, Sakyo-ku, Kyoto 606-8502, Japan}
\author{I.G.~Alekseev}\affiliation{Institute for Theoretical and Experimental
  Physics (ITEP), 117259 Moscow, Russia}
\author{A.~Bravar}\affiliation{Brookhaven National Laboratory, Upton, NY
  11973, USA}
\author{G.~Bunce}\affiliation{Brookhaven National Laboratory, Upton, NY 11973,
USA}\affiliation{RIKEN BNL Research Center, Upton, NY 11973, USA}
\author{S.~Dhawan}\affiliation{Yale University, New Haven, CT 06520, USA}
\author{R.~Gill}\affiliation{Brookhaven National Laboratory, Upton, NY 11973, USA}
\author{W.~Haeberli}\affiliation{University of Wisconsin, Madison, WI 53706, USA}
\author{O.~Jinnouchi}\thanks{Present address: KEK, Tsukuba, Ibaraki 305,
  Japan}\affiliation{RIKEN BNL Research Center, Upton, NY 11973, USA}
\author{A.~Khodinov}\affiliation{Stony Brook University, Stony Brook, NY 11794, USA}
\author{Y.~Makdisi}\affiliation{Brookhaven National Laboratory, Upton, NY
  11973, USA}
\author{A.~Nass}\thanks{Present address: University of Erlangen - Nurenberg,
  91058 Erlangen, Germany}\affiliation{Brookhaven National Laboratory, Upton, NY 11973, USA}
\author{N.~Saito}\affiliation{Kyoto University, Sakyo-ku, Kyoto 606-8502, Japan}
\author{E.J.~Stephenson}\affiliation{Indiana University Cyclotron Facility, Bloomington, IN
  47408, USA}
\author{D.N.~Svirida}\affiliation{Institute for Theoretical and Experimental
  Physics (ITEP), 117259 Moscow, Russia}
\author{T.~Wise}\affiliation{University of Wisconsin, Madison, WI 53706, USA}
\author{A.~Zelenski}\affiliation{Brookhaven National Laboratory, Upton, NY
  11973, USA}

%\collaboration{RHIC Polarimetry Group}\homepage{www4.rcf.bnl.gov/\~cnipol/}\noaffiliation

\date{\today}

\begin{abstract}

A precise measurement of the analyzing power $A_N$
in proton-proton elastic~scattering~in~the~region
of 4-momentum transfer squared
$0.001 < |t| < 0.032~({\rm GeV}/c)^2$
has been performed
using a po\-la\-rized atomic hydrogen gas jet target and the 100~GeV/$c$
RHIC proton beam.
The interference of the electromagnetic spin-flip amplitude with a
hadronic spin-nonflip amplitude is predicted to generate a
significant $A_N$ of 4--5\%, peaking at $-t \simeq 0.003~({\rm GeV}/c)^2$.
This kinematic region 
is known as the Coulomb Nuclear Interference region.
A possible hadronic spin-flip amplitude modifies this
otherwise calculable prediction.
Our data are well described by the CNI prediction with
the electromagnetic spin-flip alone
and do not support the presence of a large hadronic spin-flip amplitude.

\end{abstract}

\maketitle

%%%%%%%%%%%%%%%%%%%%%%%%%%%%%%%%%%%%%%%%%%%%%%%%%%%%%%%%%%%%%%%%%%%%%%
% Introduction
%%%%%%%%%%%%%%%%%%%%%%%%%%%%%%%%%%%%%%%%%%%%%%%%%%%%%%%%%%%%%%%%%%%%%%
Interference phenomena in hadron collisions have often
led to spectacular spin effects in the final state.
This, for instance, occurs in the elastic scattering of transversely
polarized protons at small angles, where the interference 
of a small spin-flip amplitude, that otherwise would be difficult to detect,
with a large spin-nonflip amplitude leads to a sizeable analyzing power $A_N$.
The region of low 4-momentum transfer squared $t$ is associated with long
distance phenomena
and therefore is in the domain of non-perturbative QCD.
Polarized experiments in this region allow us to explore the spin
properties of QCD at large distances.
$A_N$ is a measure of the {\it left--right} asymmetry
of the cross section in the scattering plane normal to the beam
or target polarization.

In high energy $pp$ and $pA$ elastic scattering at very low $t$,
the dominant contribution to $A_N$ comes from the interference
between the electromagnetic (Coulomb) spin-flip amplitude,
which is generated by the proton's anomalous magnetic moment,
and the hadronic (nuclear) spin-nonflip amplitudes,
which can be related to the total cross section $\sigma_{tot}$
via the optical theorem (thus Coulomb Nuclear Interference).
$A_N$ reaches a predicted maximum value of about 4--5\%
around $-t \simeq 3 \times 10^{-3}~({\rm GeV}/c)^2$
and decreases with increasing $|t|$~\cite{Kop74,But78}.
In general, the form of $A_N$ and the position of its maximum
depend on the parameters describing the hadronic amplitudes:
$\sigma_{tot}$,
the ratio $\rho$ between the real and imaginary parts of the forward
scattering amplitude, the Bethe phase shift $\delta_C$,
and the nuclear slope parameter $b$~\cite{But99}.

A potential hadronic spin-flip amplitude
interfering with the electromagnetic spin-nonflip one
introduces a deviation in shape and magnitude for $A_N$ calculated
with no hadronic spin-flip~\cite{But99}.
A measurement of $A_N$ in the CNI region, therefore,
can be a sensitive probe for the hadronic spin-flip amplitude.
This effect is measured by the ratio $r_5$ between the
single spin-flip ($\phi_5^{had}$) and
the spin-nonflip ($\phi_1^{had}$, $\phi_3^{had}$) hadronic amplitudes:
\begin{equation}
r_5 = \lim_{|t|\rightarrow 0}\frac{M_p}{\sqrt{-t}} 
       \frac{\phi_5^{had}}
            {{\rm Im}\left(\phi_1^{had} + \phi_3^{had} \right) / 2} \; ,
\end{equation}
where $M_p$ is the proton mass.
In the simplest hypothesis the reduced spin-flip amplitude
$\phi_5^{had}/\sqrt{-t}$ is assumed to be proportional
to $\phi_1^{had}$ and $\phi_1^{had} = \phi_3^{had}$~\cite{But99}.
The $t$ dependence of the hadronic spin-flip amplitude, however,
is tightly connected with the structure of hadrons at large distances
and carries information on the static properties
and on the constituent quark structure of the nucleon~\cite{Pre02}.

Within Regge phenomenology, one can probe the long standing issue
of the magnitude of the Pomeron spin-flip~\cite{Kop89,But99}.
In a diquark enhanced picture of the proton, for instance, the magnitude of the
hadronic spin-flip amplitude is associated with the diquark
separation, the smaller this separation the bigger the effect~\cite{Kop89}.
In the impact model based on the rotating matter picture
for a polarized proton the spin-orbit coupling provides a helicity-flip
amplitude~\cite{Sof79}.
Hadronic spin-flip contributions may also have their origins
in instantons~\cite{Ans93}.

A first measurement of $A_N$ at $\sqrt{s} = 19.4~{\rm GeV}$,
though much less precise, 
had been performed by the E704 experiment at Fermilab
using the 200~GeV/$c$ polarized proton beam obtained from the decay
of $\Lambda$ hyperons~\cite{E704}.
Recently, $A_N$ has been measured also at $\sqrt{s} = 200~{\rm GeV}$
by colliding the RHIC polarized proton beams~\cite{Bue05}.

%%%%%%%%%%%%%%%%%%%%%%%%%%%%%%%%%%%%%%%%%%%%%%%%%%%%%%%
%%%   E X P E R I M E N T A L   S E T U P           %%%
%%%%%%%%%%%%%%%%%%%%%%%%%%%%%%%%%%%%%%%%%%%%%%%%%%%%%%%

In this Letter we report on a precise measurement of 
the analyzing power $A_N$
in the CNI region of $0.001 < |t| < 0.032~({\rm GeV}/c)^2$
at $\sqrt{s} = 13.7~{\rm GeV}$
using an internal polarized atomic hydrogen gas jet target and
the 100~GeV/$c$ RHIC proton beam.
The RHIC collider accelerates transversely polarized protons
in bunches of opposite polarization~\cite{Ale03}.
By averaging over the bunch polarizations and several accelerator fills,
one obtains an unpolarized proton beam.

%%%%%%%%%%
\begin{figure}[t]
\begin{center}
\vspace*{-12mm}
\hspace*{-13mm}
\includegraphics[scale=0.40, keepaspectratio]{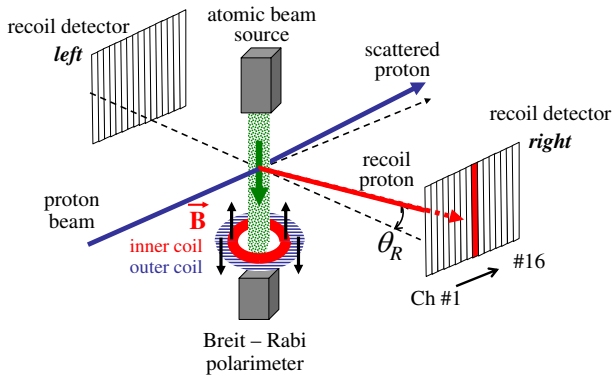}
\vspace*{-15mm}
\caption{Schematic layout of this experiment.
}
\end{center}
\label{figure:setup}
\vspace*{-5mm}
\end{figure}
%%%%%%%%%%

Figure~1 shows the schematic layout of this experiment,
located at the 12 o'clock interaction point of RHIC.
The polarized atomic hydrogen beam crossed the RHIC proton beams from above.
The two RHIC beams were radially displaced by about 10 mm, so that only the
beam circulating clockwise (the {\it blue} beam)
interacted with the jet target.
The polarization of the atomic beam was directed vertically.
In the CNI region at high energies recoil protons from $pp$ elastic scattering
emerge close to $90^\circ$ with respect to the incident beam direction.
The scattered beam protons did not exit the beam pipe
and were not detected.
In the covered $t$ region, however, the elastic process is fully constrained
by the recoil particle alone.

The polarized hydrogen jet is produced
by an atomic beam source
in which molecular hydrogen is dissociated by a radio frequency (RF) discharge.
Hydrogen atoms emerge through a 2~mm diameter nozzle cooled to 70~K
and enter a set of sextupole magnets
that spin separate and focus the atomic beam according to the electron spin.
Nuclear polarization of the atoms is obtained using two RF transitions
that induce spin-flips in the hydrogen atoms.
To avoid depolarization of the atoms
a set of Helmholtz coils around the interaction point provided
a very uniform vertical magnetic holding field (0.12~T).

The target polarization was constantly monitored with a Breit-Rabi
polarimeter located below the interaction point. 
The net proton polarization was $0.958 \pm 0.001$.
A measured $(3.5 \pm 2.0)\%$ contamination of the atomic beam
by hydrogen atoms bound into molecules
resulted in a small dilution of the target polarization.
Taking into account this dilution,
the target polarization was $P_T = 0.924 \pm 0.018$.
The proton polarization was reversed every 600~s
by turning on one or the other of two RF coils.
The efficiency of the spin-flip transitions was above 99\%.
Most systematic effects associated with the 
spin-asymmetry extraction thus cancel.
The atomic jet achieved a polarized beam intensity of
$1.2 \times 10^{17}$~H~atoms/s at a speed of $(1560 \pm 50)~{\rm m/s}$,
which is the highest intensity recorded to date.
At the interaction point the target profile is nearly gaussian 
with a FWHM of 6.5~mm.
The areal density of the target is
$(1.3 \pm 0.2) \times 10^{12}~{\rm H/cm}^2$.
For more details see~\cite{Zel05}.

%%%%%%%%%%%%%%%%%%%%%%%%%%%%%%%%%%%%%%%%%%%%%%%%%%%%%
%%%     R E C O I L   S P E C T R O M E T E R     %%%
%%%%%%%%%%%%%%%%%%%%%%%%%%%%%%%%%%%%%%%%%%%%%%%%%%%%%

The recoil protons were detected using an array of silicon detectors
located to the left and to the right of the beam 
at a distance $D \simeq 80~{\rm cm}$.
On each side, the detectors cover an aziumuthal angle of $15^\circ$
centered on the horizontal mid-plane.
Since the momenta of the recoil protons are very low,
a second set of Helmholtz coils, coaxial to the first one with
the current circulating in the opposite direction,
was used to cancel the deflection of the recoil proton trajectory
induced by the inner coils.
The resulting total magnetic field integral $\int B dl$
seen by the recoil protons is nearly zero,
and the deviation from the original trajectory
was less than 3~mm
for the lowest momentum detected protons,
leading to an almost identical acceptance for the left and right
detectors.

Each array consisted of 3 silicon detectors segmented horizontally,
$80 \times 50~{\rm mm}^2$ in size, with a $4.4~{\rm mm}$ read out pitch
for a total of 16 channels per detector.
The detectors were $\sim 400~\mu{\rm m}$ thick.
Recoil protons with kinetic energies $T_R$ up to 7~MeV were
fully absorbed. 
More energetic protons punched through the detectors,
depositing only a fraction of their energy.
The energy calibration of the silicon detectors was performed using
two $\alpha$ sources of different energies
($^{148}{\rm Gd}$ and $^{241}{\rm Am}$),
which allowed us
also to estimate the thickness of the detector's entrance window
($\sim 2~\mu{\rm m}$) and correct for it.
The detectors were read out with waveform digitizers (WFD)
that performed simultanously the function of peak sensing ADC's and
constant fraction discriminators,
and provided a deadtime free data acquisition system.

The recoil detectors provided energy ($T_R$),
recoil polar angle ($\vartheta_R$), and time of flight (ToF)
measurements for the recoil particles.
The 4-momentum transfer squared is given by $-t = 2 M_p T_R$.
The ToF is measured with respect to the bunch crossing
given by the accelerator RF clock.
Typical resolutions were
$\Delta T_R \leq 60~{\rm keV}$, $\Delta\vartheta_R \simeq 3.8~{\rm mrad}$,
and $\Delta{\rm ToF} \simeq 3~{\rm ns}$.
The angular resolution receives contributions from the
spatial resolution of the detector ($\simeq 1.6$~mrad) 
and the jet target profile.
The ToF resolution comes from the intrinsic time resolution
of the detectors ($\leq 2~{\rm ns}$) and the length of
the RHIC beam bunches ($\sigma \simeq 1.5~{\rm ns}$).

%%%%%%%%%%
\begin{figure}[t]
\vspace*{-3mm}
\begin{center}
\includegraphics[width=1.0\linewidth]{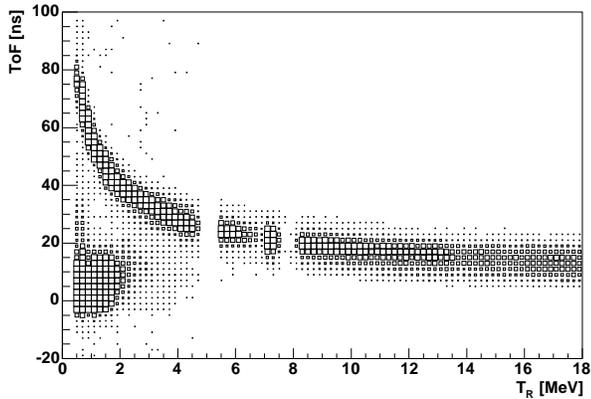}
\vspace*{-2mm}
\caption{ToF vs $T_R$ scatter plot for all recorded events.
The locus is populated by recoil protons,
while the bulk of events on the bottom left is due to
prompt and beam halo events.
The {\it empty} vertical bands are populated by the calibration
$\alpha$ sources and have been removed from the plot.}
\end{center}
\label{figure:ToF}
\vspace*{-7mm}
\end{figure}
%%%%%%%%%%

%%%%%%%%%%%%%%%%%%%%%%%%%%%%%%%%%%%%%%%%%%%%%
%%%     E V E N T   S E L E C T I O N     %%%
%%%%%%%%%%%%%%%%%%%%%%%%%%%%%%%%%%%%%%%%%%%%%

Recoil protons were identified using the
non-relativistic relation
$T_R = \frac{1}{2} M_p (D/{\rm ToF})^2$ shown in Figure~2.
Recoil protons of given $T_R$ could be clearly separated
from prompt particles on the ToF basis.
Below 7~MeV recoil protons are fully absorbed in the detectors.
Above 7~MeV, $T_R$ is corrected for punch-through
using the detector thickness and the energy loss in silicon~\cite{Berxx}.
Events were selected in a ToF interval of $\pm 8$~ns
around the expected ToF value for recoil protons of a given $T_R$.

%%%%%%%%%%
\begin{figure}[b]
\vspace*{-4mm}
\begin{center}
\includegraphics[width=1.0\linewidth]{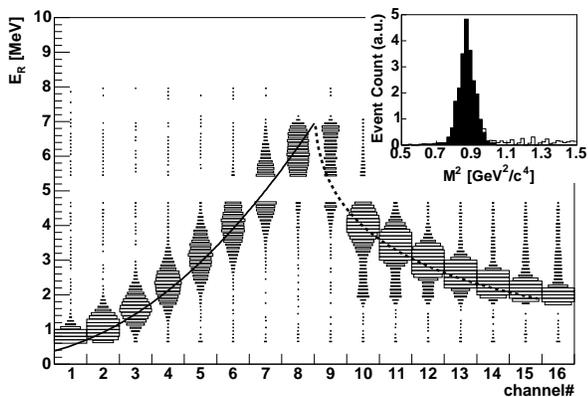}
\vspace*{-2mm}
\caption{$E_R$ vs. $\vartheta_R$ correlation:
$\vartheta_R$ is measured by the detector channel number in
$\simeq 5.5~{\rm mrad}$ steps.
The inset shows the missing mass squared $M_X^2$.
for $1 \leq T_R \leq 3~{\rm MeV}$.
}
\end{center}
\label{figure:thetaR}
\vspace*{-5mm}
\end{figure}
%%%%%%%%%%

$pp$ elastic scattering events are identified on the basis of the
$\vartheta_R$ -- $T_R$ relation
$T_R \simeq 2 M_p \vartheta_R^2$.
Figure~3 shows the measured $E_R$ -- $\vartheta_R$ correlation
for ToF selected events.
In the scatter plot
$\vartheta_R$ is measured by the detector channel number in 
$\simeq~5.5$ mrad steps.
$E_R$ is the energy deposited by the recoil protons in the detectors.
The locus on the left in Figure~3 is generated by fully absorbed protons,
while the locus on the right is due to punch through protons.
The full line is the kinematic expectation
for $pp$ elastic scattering for fully absorbed protons.
The dashed line shows the expected energy deposited in the detector
by more energetic protons for $T_R > 7~{\rm MeV}$.

For each $T_R$ bin (see Table~1) $pp$ elastic events were selected
in 3 adjacent detector channels centered around the expected
$\vartheta_R$ angle. 
On the basis of the measured $\vartheta_R$ and $T_R$
the mass of the undetected scattered beam particle
(the missing mass $M_X$)
can be reconstructed.
For $pp$ elastic scattering $M_X = M_p$.
The channel for diffractive dissociation opens at
$M_X > M_p + M_\pi = 1.08~{\rm GeV}/c^2$.
The selected $pp$ elastic events
are well separated from the inelastic threshold for $T_R < 8~{\rm MeV}$
(see inset).
The contamination from inelastic channels for larger $T_R$ was estimated
to be less than 0.5\%.

%%%%%%%%%%%%%%%%%%%%%%%%%%%%%%%%%%%%%%%
%%%          R E S U L T S          %%%
%%%%%%%%%%%%%%%%%%%%%%%%%%%%%%%%%%%%%%%

$A_N$ was extracted from the geometrical mean of spin sorted event
yields~\cite{Ohl73}:
\begin{equation}
A_N = - \frac{1}{P_T}
            \frac{\sqrt{N^\uparrow_L \cdot N^\downarrow_R}-
            \sqrt{N^\uparrow_R \cdot N^\downarrow_L}}
           {\sqrt{N^\uparrow_L \cdot N^\downarrow_R}+
            \sqrt{N^\uparrow_R \cdot N^\downarrow_L}}
\end{equation}
where $N^{\uparrow(\downarrow)}_{L(R)}$ is the number of selected $pp$
elastic scattering
events detected on the left(right) of the beam,
and the arrows give the direction of the target polarization.
In this expression flux factors, acceptances, and efficiencies
factor out and appear only as third order corrections.
Data were gathered in 14~$T_R$ bins (see Table~I).

The level of background under the signal was estimated from
empty target runs with and without beam,
and sidebands in the strip distributions for fixed recoil energy bins.
These backgrounds were flat.
The background included $\alpha$ source background,
beam scraping, and beam scattering from residual target gas.
The overall background level was dependent on the $T_R$ bin:
it increased from 6\% to 9\% for the lowest $T_R$ bins (0.6 to 7.4~MeV)
and $\simeq 10\%$ for the punch-through bins (8.0 to 17.0~MeV).
The final $A_N$ values were corrected for $\alpha$ source background
and beam gas scattering.
The background originating from the unpolarized residual target gas
and the target tails
has been already accounted for as a dilution of the target polarization.

The major sources of systematic uncertainties come from the
error on the target polarization
giving an overall 2.0\% normalization uncertainty,
the false and acceptance asymmetries,
event selection criteria, and the background.
The background uncertainty from the residual target gas
is included in the target polarization uncertainty,
which is from two independent measurements.

Figure~4 displays the analyzing power $A_N$ as a function of $t$
in the range $0.001 < |t| < 0.032~({\rm GeV}/c)^2$
at $\sqrt{s} = 13.7~{\rm GeV}$.
These data are summarized in Table~1.

%%%%%%%%%%
\begin{figure}[t]
\vspace*{-2mm}
\begin{center}
\includegraphics[width=1.0\linewidth]{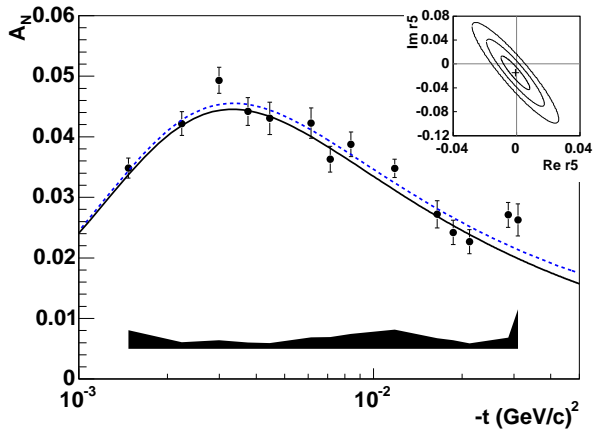}
\caption{$A_N$ as a function of $t$ for $pp^\uparrow \rightarrow pp$
at $\sqrt{s} = 13.7~{\rm GeV}$.
The errors on the data points are statistical.
The lower band represents the total systematic error.
The prediction for $A_N$ with the electromagnetic spin-flip 
only is superimposed to the data (solid lane).
The dashed line is a fit to the data allowing for a hadronic spin-flip
contribution to $A_N$.
Inset: $r_5$ with the 1-, 2-, and 3-$\sigma$ confidence contours.
}
\end{center}
\label{figure:AN}
\vspace*{-4mm}
\end{figure}
%%%%%%%%%

The $A_N$ data are compared to the CNI prediction
with no hadronic spin-flip~\cite{But99}
using for the hadronic amplitudes
$\sigma_{tot} = 38.4~{\rm mb}$, $\rho = -0.08$, $\delta_C = 0.02$,
$b = 12~({\rm GeV}/c)^{-2}$~\cite{Blo85}.
The major uncertainty in the CNI prediction comes from the
parametrization of the hadronic amplitudes
and the approximate knowledge of the $\rho$ parameter.
The $\chi^2$ is 13.4 for 14 degrees of freedom.

The $A_N$ data were also fitted with the CNI prediction allowing for
a hadronic spin-flip contribution (Eq.~1, dashed line in
Figure~4).
The quality of the fit is similar to the case with no
hadronic spin-flip ($\chi^2 = 11.1 / 12$~d.o.f.).
The values obtained for $r_5$ are
${\rm Re}~r_5 = -0.0008 \pm 0.0091$
and ${\rm Im}~r_5 = -0.015 \pm 0.029$,
and the correlation parameter between ${\rm Re}~r_5$ and ${\rm Im}~r_5$
is $-0.92$.
The results of the $r_5$ fit are shown as inset in Figure~4.

These data are consistent with no hadronic spin-flip
and do not support the presence of a large hadronic spin-flip
amplitude at this energy.
$A_N$ data from proton-carbon elastic scattering over a similar
kinematic range at the same~\cite{Jin04} and lower~\cite{Toj02} energies,
on the contrary, deviate substantially from the simple CNI prediction and
require a substantial hadronic spin-flip contribution.

%%%%%%%%%%%%%%%%%%%%%%%%%%%%%
%%%     S U M M A R Y     %%%
%%%%%%%%%%%%%%%%%%%%%%%%%%%%%

In summary, we have measured $A_N$ for $pp$ elastic
scattering at $\sqrt{s} = 13.7~{\rm GeV}$ over the $t$ range of
$0.001 < |t| < 0.032~({\rm GeV}/c)^2$.
Our data are well described by the CNI prediction,
in which $A_N$ is generated by the pro\-ton's anomalous magnetic moment alone,
and do not support the presence of a large hadronic spin-flip
contribution.
Measurements of similar precision at higher as well as lower energies,
and using different nuclear targets
are required to fully disentangle the role of the hadronic spin-flip
amplitudes, their energy dependence and the different behavior
between proton and nuclear targets.

%%%%%%%%%%%%%%%%%%%%%%%%%%%%%%%%%%%%%%%%%%%%%%%
%%%     A C K N O W L E D G E M E N T S     %%%
%%%%%%%%%%%%%%%%%%%%%%%%%%%%%%%%%%%%%%%%%%%%%%%

We would like to thank Collider Accelerator Department and the Instrumentation Division at BNL for
their work on the silicon detectors and electronics.
This work is performed under the auspices of U.S. DOE contract
Nos. DE-AC02-98CH10886 and W-31-109-ENG-38, DOE grant No. DE-FG02-88ER40438,
NSF grant PHY-0100348, and with support from RIKEN, Japan.
\\

\newpage

%%%%%%%%%%%%%%%%%%%%%%
\begin{table}
\begin{tabular}{c|c|c|c}
\hline
\hline
$T_R$ & $-\langle t \rangle$ & $A_N$ & $\delta A_N$ \\
(MeV) & $({\rm GeV}^2/c^2)$  &       & (stat. $\pm$ sys. $\pm$ norm.)  \\
\hline
0.6 -- 1.0   &  0.0015 &  $\; 0.0348 \;$ &   $0.0017 \pm 0.0030 \pm 0.0007$ \\
1.0 -- 1.4   &  0.0022 &  $\; 0.0422 \;$ &   $0.0020 \pm 0.0007 \pm 0.0008$ \\
1.4 -- 1.8   &  0.0030 &  $\; 0.0493 \;$ &   $0.0022 \pm 0.0010 \pm 0.0010$ \\
1.8 -- 2.2   &  0.0037 &  $\; 0.0442 \;$ &   $0.0023 \pm 0.0006 \pm 0.0009$ \\
2.2 -- 2.5   &  0.0044 &  $\; 0.0430 \;$ &   $0.0027 \pm 0.0004 \pm 0.0008$ \\
3.0 -- 3.5   &  0.0061 &  $\; 0.0423 \;$ &   $0.0025 \pm 0.0017 \pm 0.0008$ \\
3.5 -- 4.2   &  0.0071 &  $\; 0.0363 \;$ &   $0.0021 \pm 0.0018 \pm 0.0007$ \\
4.2 -- 4.7   &  0.0084 &  $\; 0.0388 \;$ &   $0.0020 \pm 0.0023 \pm 0.0008$ \\
5.7 -- 7.4   &  0.0118 &  $\; 0.0348 \;$ &   $0.0015 \pm 0.0031 \pm 0.0007$ \\
8.0 -- 9.3   &  0.0165 &  $\; 0.0272 \;$ &   $0.0023 \pm 0.0016 \pm 0.0005$ \\
9.3 -- 10.6  &  0.0187 &  $\; 0.0242 \;$ &   $0.0020 \pm 0.0013 \pm 0.0005$ \\
10.6 -- 12.0 &  0.0212 &  $\; 0.0227 \;$ &   $0.0020 \pm 0.0008 \pm 0.0004$ \\
14.5 -- 16.0 &  0.0287 &  $\; 0.0271 \;$ &   $0.0021 \pm 0.0018 \pm 0.0005$ \\
16.0 -- 17.0 &  0.0309 &  $\; 0.0263 \;$ &   $0.0027 \pm 0.0065 \pm 0.0005$ \\
\hline
\hline
\end{tabular}
\caption{$A_N$ as a function of $t$ in 14 $T_R$ bins.
The first error is the statistical one, followed by the systematic error,
and the normalization error on $P_T$.}
\label{table:AN}
\end{table}
%%%%%%%%%%%%%%%%%%%%%%

%%%%%%%%%%%%%%%%%%%%%%

\end{document}